\documentclass[12pt,a4paper]{article}
\pdfoutput=1
\usepackage{jheppub}
\usepackage{amsfonts, amsthm}
\usepackage[english]{babel}
\usepackage[utf8]{inputenc}
\usepackage{mathrsfs}
\usepackage{amssymb}
\usepackage{color}
\hypersetup{unicode}

\newcommand{\eq}{\begin{equation}}
\newcommand{\feq}{\end{equation}}
\newcommand{\beq}{\begin{equation}}
\newcommand{\eeq}{\end{equation}}
\newcommand{\eqNN}{\begin{equation*}}
\newcommand{\feqNN}{\end{equation*}}
\font\mybb=msbm10 at 12pt
\def\bb#1{\hbox{\mybb#1}}

\def\bR {\bb{R}}

\newcommand{\CC}{\mathbb{C}}
\newcommand{\RR}{\mathbb{R}}
\newcommand{\ZZ}{\mathbb{Z}}

\newcommand{\D}{{\rm d}}

\title{Supersymmetric black holes with spiky horizons}

\author[1,2]{Federico Faedo,}
\author[2,3]{Silke Klemm}
\author[2,3]{and Adriano Vigan\`o}

\affiliation[1]{Dipartimento di Matematica, Universit\`a di Torino, \\
Via Carlo Alberto 10, I-10123 Torino}
\affiliation[2]{INFN, Sezione di Milano, \\
Via Celoria 16, I-20133 Milano}
\affiliation[3]{Dipartimento di Fisica, Universit\`a di Milano, \\
Via Celoria 16, I-20133 Milano}

\emailAdd{federicomichele.faedo@unito.it}
\emailAdd{silke.klemm@mi.infn.it}
\emailAdd{adriano.vigano@unimi.it}
\preprint{IFUM-1088-FT}

\abstract{We use the recipe of \cite{Klemm:2010mc} to find half-BPS near-horizon geometries in the t$^3$ 
model of $N=2$, $D=4$ gauged supergravity, and explicitely construct some new examples. Among these
are black holes with noncompact horizons, but also with spherical horizons that have conical singularities 
(spikes) at one of the two poles. A particular family of them is extended to the full black hole geometry.
Applying a double-Wick rotation to the near-horizon region, we obtain solutions with NUT charge
that asymptote to curved domain walls with AdS$_3$ world volume.
These new solutions may provide interesting testgrounds to address fundamental questions related
to quantum gravity and holography.}


\begin{document}

\maketitle

\flushbottom

\section{Introduction and summary of results}

Since the formulation of the AdS/CFT correspondence, asymptotically AdS black holes have been
attracting much interest, partly due to their application to entropy calculations through the counting of black hole microstates (see~\cite{Benini:2015eyy,Benini:2016hjo,Benini:2016rke} for examples in $N=2$, $D=4$ Fayet--Iliopoulos (FI)-gauged supergravity). In this context gauged supergravity theories play an important role, since the gauging of (part of) the R-symmetry generically leads to a scalar potential which, in a vacuum configuration, behaves effectively as a cosmological constant and allows for non-asymptotically flat solutions that often (but not always) approach AdS at infinity.

It has been known for many years that a negative cosmological constant allows for black holes with non-spherical horizons, cf.~e.g.~\cite{Vanzo:1997gw} and references therein. Gauged supergravity is thus an interesting terrain to study these kinds of black holes. Indeed, the scalar potential typically violates the assumptions that go into uniqueness theorems, so that one can have horizons that are flat, hyperbolic or compact Riemann surfaces of any genus~\cite{Caldarelli:1998hg,Cacciatori:2009iz,DallAgata:2010ejj}.

In these contexts, cubic models are of special interest. Like in the ungauged case, some cubic models in gauged supergravity can be embedded into higher-dimensional theories describing the low energy limit of some string theory. In particular, the FI-gauged stu model, containing three vector multiplets, can be obtained as a consistent truncation of eleven-dimensional supergravity compactified on a 7-sphere~\cite{Duff:1999gh,Cvetic:1999xp}. A special subcase is the so-called t$^3$ model, where the three vector multiplets are identified and which will be considered in this work.

Due to its importance, the gauged stu model has been extensively studied in the past. The first example appeared in~\cite{Duff:1999gh}, where the four-dimensional $N=2$ theory was obtained as a truncation of $N=8$ gauged supergravity, and static nonextremal black holes carrying electric or magnetic charges were presented. Subsequently, static and BPS generalizations were constructed in~\cite{Cacciatori:2009iz,Sabra:1999ux,Chamseddine:2000bk,Hristov:2010ri,Katmadas:2014faa, Halmagyi:2014qza}, while nonextremal solutions to the t$^3$ and stu models were found in~\cite{Klemm:2012yg,Klemm:2012vm}. Rotation was added in~\cite{Klemm:2011xw,Gnecchi:2013mja}, and more recently in~\cite{Hristov:2018spe,Daniele:2019rpr,Hristov:2019mqp}. Moreover, in~\cite{Hristov:2018spe} a general analysis of the possible asymptotic behaviours was performed by means of the real formulation of special geometry, in terms of a specific quartic form $I_4$ invariant under symplectic transformations. This provides a way to treat symplectically related prepotentials on the same footing, including theories which do not admit a prepotential. Although in the ungauged case all these theories lead to the same physics, the gauging breaks symplectic invariance, producing systems with different behaviours for the same gauge couplings, e.g.~an AdS boundary at infinity or something completely different.

In this paper we will focus on the t$^3$ model with prepotential $F=-(X^1)^3/X^0$ and electric gauging.
Using the classification of half-supersymmetric backgrounds presented in~\cite{Klemm:2010mc}, we will be 
able to obtain three classes of near-horizon solutions. Unfortunately, none of their possible full black
hole extensions is expected to be asymptotically AdS$_4$ due to the absence of critical points of the
scalar potential. Specifically, in section \ref{sec:sugra} we briefly review $N=2$, $D=4$ FI-gauged 
supergravity and the classifications of 1/4- and 1/2-BPS backgrounds given respectively in~\cite{Cacciatori:2008ek} and~\cite{Klemm:2010mc}. In section~\ref{sec:scalars} we construct three new
classes of near-horizon geometries, one of which turns out to be a generalization of the solution derived
in section 4 of~\cite{Daniele:2019rpr}. Sections~\ref{sec:first-solution} and~\ref{sec:second-solution} are devoted to a deeper analysis of two of the new solutions, that can either have noncompact horizons, or 
spherical horizons with conical singularities (spikes) at one of the two poles. In~\ref{sec:third-solution} a
full black hole extension of the third one is presented. Finally, applying a double-Wick rotation (which 
amounts to an analytical continuation of the coordinates) to the
first two solutions, we obtain configurations with NUT charge that asymptote to curved domain walls
with AdS$_3$ world volume, where AdS$_3$ appears as a Hopf-like fibration over H$^2$.

\section{$N=2$, $D=4$ FI-gauged supergravity}
\label{sec:sugra}

\subsection{The theory and BPS equations}
\label{subsec:BPSeqns}

We consider $N=2$, $D=4$ gauged supergravity coupled to $n$ abelian
vector multiplets \cite{Andrianopoli:1996cm}\footnote{Throughout this paper,
we use the notations and conventions of \cite{Freedman:2012zz}.}.
Apart from the vierbein $e^a_{\mu}$, the bosonic field content includes the
vectors $A^I_{\mu}$ enumerated by $I=0,\ldots,n$, and the complex scalars
$z^{\alpha}$ where $\alpha=1,\ldots,n$. These scalars parametrize
a special K\"ahler manifold, i.e., an $n$-dimensional
Hodge--K\"ahler manifold that is the base of a symplectic bundle, with the
covariantly holomorphic sections
\begin{equation}
{\cal V} = \left(\begin{array}{c} X^I \\ F_I\end{array}\right)\,, \qquad
{\cal D}_{\bar\alpha}{\cal V} = \partial_{\bar\alpha}{\cal V}-\frac 12
(\partial_{\bar\alpha}{\cal K}){\cal V}=0\,, \label{sympl-vec}
\end{equation}
where ${\cal K}$ is the K\"ahler potential and ${\cal D}$ denotes the
K\"ahler-covariant derivative. ${\cal V}$ obeys the symplectic constraint
\begin{equation}
\langle {\cal V},\bar{\cal V}\rangle = X^I\bar F_I-F_I\bar X^I=i\,. \label{sympconst}
\end{equation}
To solve this condition, one defines
\begin{equation}
{\cal V}=e^{{\cal K}(z,\bar z)/2}v(z)\,,
\end{equation}
where $v(z)$ is a holomorphic symplectic vector,
\begin{equation}
v(z) = \left(\begin{array}{c} Z^I(z) \\ \frac{\partial}{\partial Z^I}F(Z)
\end{array}\right)\,.
\end{equation}
$F$ is a homogeneous function of degree two, called the prepotential,
whose existence is assumed to obtain the last expression.
The K\"ahler potential is then
\begin{equation}
e^{-{\cal K}(z,\bar z)} = -i\langle v,\bar v\rangle\,.
\end{equation}
The matrix ${\cal N}_{IJ}$ determining the coupling between the scalars
$z^{\alpha}$ and the vectors $A^I_{\mu}$ is defined by the relations
\begin{equation}\label{defN}
F_I = {\cal N}_{IJ}X^J\,, \qquad {\cal D}_{\bar\alpha}\bar F_I = {\cal N}_{IJ}
{\cal D}_{\bar\alpha}\bar X^J\,.
\end{equation}
The bosonic action reads
\begin{equation}
\label{action}
\begin{split}
e^{-1}{\cal L}_{\text{bos}} &= \frac 12R + \frac 14(\text{Im}\,
{\cal N})_{IJ}F^I_{\mu\nu}F^{J\mu\nu} - \frac 18(\text{Re}\,{\cal N})_{IJ}\,e^{-1}
\epsilon^{\mu\nu\rho\sigma}F^I_{\mu\nu}F^J_{\rho\sigma} \\
& \quad -g_{\alpha\bar\beta}\partial_{\mu}z^{\alpha}\partial^{\mu}\bar z^{\bar\beta}
- V \, ,
\end{split}
\end{equation}
with the scalar potential
\eq
V = -2g^2\xi_I\xi_J[(\text{Im}\,{\cal N})^{-1|IJ}+8\bar X^IX^J]\,, \label{scal-pot}
\feq
that results from U$(1)$ Fayet--Iliopoulos gauging. Here, $g$ denotes the
gauge coupling and the $\xi_I$ are FI constants. In what follows, we define
$g_I\equiv g\xi_I$.

The most general timelike supersymmetric background of the theory described
above was constructed in \cite{Cacciatori:2008ek}, and is given by
\eq
ds^2 = -4|b|^2(dt+\sigma)^2 + |b|^{-2}(dz^2+e^{2\Phi} dw d\bar w)\ , \label{gen-metr}
\feq
where the complex function $b(z,w,\bar w)$, the real function $\Phi(z,w,\bar w)$
and the one-form $\sigma=\sigma_wdw+\sigma_{\bar w}d\bar w$, together with the
symplectic section~\eqref{sympl-vec}\footnote{Note that also $\sigma$ and
$\cal V$ are independent of $t$.} are determined by the equations
\eq
\partial_z\Phi = 2ig_I\left(\frac{{\bar X}^I}b-\frac{X^I}{\bar b}\right)\ ,
\label{dzPhi}
\feq
\eq
\begin{split}
& 4\partial\bar\partial\left(\frac{X^I}{\bar b}-\frac{\bar X^I}b\right) + \partial_z\left[e^{2\Phi}\partial_z
\left(\frac{X^I}{\bar b}-\frac{\bar X^I}b\right)\right]  \label{bianchi} \\
& \qquad -2ig_J\partial_z\left\{e^{2\Phi}\left[|b|^{-2}(\text{Im}\,{\cal N})^{-1|IJ}
+ 2\left(\frac{X^I}{\bar b}+\frac{\bar X^I}b\right)\left(\frac{X^J}{\bar b}+\frac{\bar X^J}b\right)\right]\right\}= 0\,,
\end{split}
\feq
\eq
\label{maxwell}
\begin{split}
& 4\partial\bar\partial\left(\frac{F_I}{\bar b}-\frac{\bar F_I}b\right) + \partial_z\left[e^{2\Phi}\partial_z
\left(\frac{F_I}{\bar b}-\frac{\bar F_I}b\right)\right]  \\
& \qquad -2ig_J\partial_z\left\{e^{2\Phi}\left[|b|^{-2}\text{Re}\,{\cal N}_{IL}(\text{Im}\,{\cal N})^{-1|JL}
+ 2\left(\frac{F_I}{\bar b}+\frac{\bar F_I}b\right)\left(\frac{X^J}{\bar b}+\frac{\bar X^J}b\right)\right]\right\} \\
& \qquad -8ig_I e^{2\Phi}\left[\langle {\cal I}\,,\partial_z {\cal I}\rangle-\frac{g_J}{|b|^2}\left(\frac{X^J}{\bar b}
+\frac{\bar X^J}b\right)\right] = 0\,, 
\end{split}
\feq
\begin{equation}
2\partial\bar\partial\Phi=e^{2\Phi}\left[ig_I\partial_z\left(\frac{X^I}{\bar b}-\frac{\bar X^I}b\right)
+\frac2{|b|^2}g_Ig_J(\text{Im}\,{\cal N})^{-1|IJ}+4\left(\frac{g_I X^I}{\bar b}+\frac{g_I \bar X^I}b
\right)^2\right]\,, \label{Delta-Phi}
\end{equation}
\begin{equation}
d\sigma + 2\,\star^{(3)}\!\langle{\cal I}\,,d{\cal I}\rangle - \frac i{|b|^2}g_I\left(\frac{\bar X^I}b
+\frac{X^I}{\bar b}\right)e^{2\Phi}dw\wedge d\bar w=0\,. \label{dsigma}
\end{equation}
Here $\star^{(3)}$ is the Hodge star on the three-dimensional base with metric\footnote{Whereas
in the ungauged case, this base space is flat and thus has trivial holonomy, here we have U(1)
holonomy with torsion \cite{Cacciatori:2008ek}.}
\eq
ds_3^2 = dz^2+e^{2\Phi}dwd\bar w\,, \label{metr-base}
\feq
and we defined $\partial=\partial_w$, $\bar\partial=\partial_{\bar w}$, as well as
\begin{equation}
{\cal I} = \text{Im}\left({\cal V}/\bar b\right)\,, \qquad {\cal R} = \text{Re}\left({\cal V}/\bar b\right)\,.
\end{equation}
Note that the eqns.~\eqref{dzPhi}-\eqref{Delta-Phi} can be written compactly in the symplectically
covariant form
\begin{equation}
\partial_z\Phi = 4\langle{\cal I},{\cal G}\rangle\,, \label{partialzPhi}
\end{equation}
\begin{equation}
\begin{split}
\Delta{\cal I} &+ 2 e^{-2\Phi}\partial_z\left\{e^{2\Phi}\left[\langle{\cal R},{\cal I}\rangle\Omega{\cal M}
{\cal G} - 4{\cal R}\langle{\cal R},{\cal G}\rangle\right]\right\} \\
&- 4{\cal G}\left[\langle{\cal I},\partial_z
{\cal I}\rangle + 4\langle{\cal R},{\cal I}\rangle\langle{\cal R},{\cal G}\rangle\right] = 0\,, 
\end{split} \label{Delta-I}
\end{equation}
\begin{equation}
\Delta\Phi = -8\langle{\cal R},{\cal I}\rangle\left[{\cal G}^t{\cal M}{\cal G} + 8|{\cal L}|^2\right]
= 4\langle{\cal R},{\cal I}\rangle V\,, \label{Delta-Phi-cov}
\end{equation}
where ${\cal G}=(g^I,g_I)^t$ represents the symplectic vector of gauge couplings\footnote{In the case
considered here with electric gaugings only, one has $g^I=0$.}, ${\cal L}=\langle{\cal V},{\cal G}\rangle$,
$\Delta$ denotes the covariant Laplacian associated to the base space metric \eqref{metr-base}, and $V$
in \eqref{Delta-Phi-cov} is the scalar potential \eqref{scal-pot}. Moreover,
\eq
\label{omega}
\Omega = \left(\begin{array}{cc} 0 & 1 \\ -1 & 0\end{array}\right)\,, \qquad {\cal M} =
\left(\begin{array}{cc}\text{Im}\,{\cal N} + \text{Re}\,{\cal N}(\text{Im}\,{\cal N})^{-1}\text{Re}\,{\cal N} & 
-\text{Re}\,{\cal N}(\text{Im}\,{\cal N})^{-1} \\ -(\text{Im}\,{\cal N})^{-1}\text{Re}\,{\cal N} &
(\text{Im}\,{\cal N})^{-1}\end{array}\right)\,.
\feq
Finally, \eqref{dsigma} can be rewritten as
\eq
d\sigma + \star_h\left(d\Sigma - A + \frac12\nu\Sigma\right) = 0\,, \label{gen-mon}
\feq
where the function $\Sigma$ and the one-form $\nu$ are respectively given by
\eq
\Sigma = \frac i2\ln\frac{\bar b}b\,, \qquad \nu = \frac8\Sigma\langle{\cal G},{\cal R}\rangle dz\,,
\feq
$A$ is the gauge field of the K\"ahler U$(1)$,
\eq
A_{\mu} = -\frac i2(\partial_{\alpha}{\cal K}\partial_{\mu}z^{\alpha} -
         \partial_{\bar\alpha}{\cal K}\partial_{\mu}{\bar z}^{\bar\alpha})\,,
\feq
and $\star_h$ denotes the Hodge star on the Weyl-rescaled base space metric
\eq
h_{ij}dx^i dx^j = \frac1{|b|^4}(dz^2+e^{2\Phi}dwd\bar w)\,.
\feq
\eqref{gen-mon} is the generalized monopole equation \cite{Jones:1985pla}, or more precisely
a K\"ahler-covariant generalization thereof, due to the presence of the one-form $A$.
In order to cast \eqref{dsigma} into the form \eqref{gen-mon}, one has to use the special K\"ahler
identities
\eq\label{identitiesDVV}
\langle{\cal D}_\alpha{\cal V},{\cal V}\rangle = \langle{\cal D}_\alpha{\cal V},\bar{\cal V}\rangle = 0\,.
\feq
Note that \eqref{gen-mon} is invariant under Weyl rescaling, accompanied by a gauge transformation
of $\nu$,
\begin{align}
h_{mn}\D x^m\D x^n \mapsto e^{2\psi}h_{mn}\D x^m\D x^n\,, \quad \Sigma \mapsto
e^{-\psi}\Sigma\,, \quad \nu \mapsto \nu + 2\D\psi\,, \quad A \mapsto e^{-\psi}A\,.
\end{align}
It would be very interesting to better understand the deeper origin of the conformal invariance
of \eqref{gen-mon} in the present context.

The integrability condition for \eqref{gen-mon} reads
\eq
D_i[h^{ij}\sqrt h (D_j - A_j)\Sigma] = 0\,, \label{int-cond}
\feq
with the Weyl-covariant derivative
\eq
D_i = \partial_i - \frac m2\nu_i\,,
\feq
where $m$ denotes the Weyl weight of the corresponding field\footnote{A field $\Gamma$ with Weyl
weight $m$ transforms as $\Gamma\mapsto e^{m\psi}\Gamma$ under a Weyl rescaling.}.
It is straightforward to show that \eqref{int-cond} is equivalent to
\eq
\langle{\cal I},\Delta{\cal I}\rangle + 4 e^{-2\Phi}\partial_z\left(e^{2\Phi}\langle{\cal I},{\cal R}\rangle
\langle{\cal G},{\cal R}\rangle\right) = 0\,,
\feq
which follows from \eqref{Delta-I} by taking the symplectic product with $\cal I$.
To shew this, one has to use
\eq
\frac12\left({\cal M} + i\Omega\right) = \Omega\bar{\cal V}{\cal V}\Omega +
\Omega {\cal D}_\alpha {\cal V} g^{\alpha\bar\beta} {\cal D}_{\bar\beta}\bar{\cal V}\Omega,
\feq
\eq
\langle {\cal D}_\alpha {\cal V}, {\cal D}_\beta {\cal V}\rangle = 0\,, \qquad
\langle {\cal D}_\alpha {\cal V}, {\cal D}_{\bar\beta}\bar{\cal V}\rangle = -i g_{\alpha\bar\beta}\,,
\feq
as well as \eqref{partialzPhi} and \eqref{identitiesDVV}.

Given $b$, $\Phi$, $\sigma$ and $\cal V$, the fluxes read
\eq
\begin{split}
F^I&=2(dt+\sigma)\wedge d\left[bX^I+\bar b\bar X^I\right]+|b|^{-2}dz\wedge d\bar w
\left[\bar X^I(\bar\partial\bar b+iA_{\bar w}\bar b)+({\cal D}_{\alpha}X^I)b\bar\partial z^{\alpha}
\right.  \\
&\quad -\left. X^I(\bar\partial b-iA_{\bar w}b)-({\cal D}_{\bar\alpha}\bar X^I)\bar b\bar\partial\bar z^{\bar\alpha}
\right]-|b|^{-2}dz\wedge dw\left[\bar X^I(\partial\bar b+iA_w\bar b)\right.  \\
&\quad +\left.({\cal D}_{\alpha}X^I)b\partial z^{\alpha}-X^I(\partial b-iA_w b)-({\cal D}_{\bar\alpha}\bar X^I)
\bar b\partial\bar z^{\bar\alpha}\right] \\
&\quad -\frac 12|b|^{-2}e^{2\Phi}dw\wedge d\bar w\left[\bar X^I(\partial_z\bar b+iA_z\bar b)+({\cal D}_{\alpha}
X^I)b\partial_z z^{\alpha}-X^I(\partial_z b-iA_z b) \right. \\
&\quad -\left.({\cal D}_{\bar\alpha}\bar X^I)\bar b\partial_z\bar z^{\bar\alpha}-2ig_J
(\text{Im}\,{\cal N})^{-1|IJ}\right]\,. \label{fluxes}
\end{split}
\feq

\subsection{1/2-BPS near-horizon geometries}
\label{1/2BPS}

An interesting class of half-supersymmetric backgrounds was obtained in \cite{Klemm:2010mc}.
It includes the near-horizon geometry of extremal rotating black holes. The metric and the
fluxes read respectively
\begin{equation}
\label{near-hor}
ds^2 = 4 e^{-\xi} \left(-r^2 dt^2 + \frac{dr^2}{r^2}\right) + 4(e^{-\xi} - K e^{\xi})
(d\phi + r\,dt)^2 + \frac{4 e^{-2\xi} d\xi^2}{Y^2 (e^{-\xi} - K e^{\xi})} \,,
\end{equation}
\begin{equation}
\label{fluxes-1/2BPS}
\begin{split}
F^I & = 16i \sqrt{K} \left(\frac{\bar{X} X^I}{1-iY} - \frac{X \bar{X}^I}{1+iY}\right) dt \wedge dr \\
    & \quad + \frac{8 \sqrt{K}} Y \left[\frac{2\bar{X} X^I}{1-iY} + \frac{2X \bar{X}^I}{1+iY} +\left(\mbox{Im}\,\mathcal{N}\right)^{-1|IJ} g_J\right] (d\phi + r\,dt) \wedge d\xi \,,
\end{split}
\end{equation}
where $X\equiv g_I X^I$, $K>0$ is a real integration constant and $Y$ is defined by
\eq
\label{Y}
Y^2 = 64 e^{-\xi} |X|^2 - 1\,.
\feq
The moduli fields $z^{\alpha}$ depend on the horizon coordinate $\xi$ only, and obey the flow
equation\footnote{Note that this is not a radial flow, but a flow along the horizon.}
\eq
\frac{dz^{\alpha}}{d\xi} = \frac i{2\bar X Y}(1-iY)g^{\alpha\bar\beta}{\cal D}_{\bar\beta}\bar X\ .
\label{dzdxi}
\feq
\eqref{near-hor} is of the form (3.3) of \cite{Astefanesei:2006dd}, and describes the near-horizon
geometry of extremal rotating black holes\footnote{Metrics of the type \eqref{near-hor} were discussed
for the first time in \cite{Bardeen:1999px} in the context of the extremal Kerr throat geometry.},
with isometry group $\text{SL}(2,\bR)\times\text{U}(1)$.
From \eqref{dzdxi} it is clear that the scalar fields have
a nontrivial dependence on the horizon coordinate $\xi$ unless $g_I{\cal D}_{\alpha}X^I=0$.
As was shown in \cite{Klemm:2010mc}, the solution with constant scalars is the near-horizon
limit of the supersymmetric rotating hyperbolic black holes in minimal gauged
supergravity \cite{Caldarelli:1998hg}.

Using $Y$ in place of $\xi$ as a new variable, \eqref{dzdxi} becomes
\begin{equation}
\frac{dz^\alpha}{dY} = \frac{X g^{\alpha\bar\beta}{\cal D}_{\bar\beta}\bar X}{(Y-i)\left[-\bar X X +
{\cal D}_\gamma X g^{\gamma\bar\delta}{\cal D}_{\bar\delta}\bar X\right]}\,. \label{flow-Y}
\end{equation}
This can also be rewritten in a K\"ahler-covariant form, as a differential equation for the symplectic
section $\cal V$,
\begin{equation}
D_Y {\cal V} = \frac{X {\cal D}_\alpha {\cal V }g^{\alpha\bar\beta}{\cal D}_{\bar\beta}\bar X}{(Y-i)
\left[-\bar X X + {\cal D}_\gamma X g^{\gamma\bar\delta}{\cal D}_{\bar\delta}\bar X\right]}\,,
\end{equation}
where
\begin{equation}
D_Y \equiv \frac d{dY} + i A_Y
\end{equation}
denotes the K\"ahler-covariant derivative.

\subsection{The $\text{t}^3$ and square root model}
\label{t3-model}

The specific t$^3$ model under investigation is defined by the prepotential
\begin{equation}
F = -\frac{{(X^1)}^3}{X^0}\,. \label{F-cubic}
\end{equation}
As mentioned before, we work in a pure electric gauging, i.e.~$\mathcal{G}=(0,0,g_0,g_1)^t$, and choose
the parametrization
\begin{equation}
Z^0 = 1\,, \qquad Z^1 = i\tau\,.
\end{equation}
Then, the holomorphic symplectic vector, the K\"ahler potential, the scalar metric and the scalar potential
are respectively given by
\begin{equation}
v = (1, i\tau, -i\tau^3, 3\tau^2)^t\,, \qquad e^{-\mathcal{K}} = (\tau + \bar\tau)^3\,, \label{vKt^3}
\end{equation}
and
\begin{equation}
g_{\tau\bar\tau} = \frac3{(\tau + \bar\tau)^2}\,, \qquad V = -\frac{8 g_1^2}{3(\tau + \bar\tau)}\,.
\label{scalar-pot-t^3}
\end{equation}
Note that $V$ is of Liouville-type, and has thus no critical points, so the theory does not admit AdS$_4$ 
vacua with constant moduli.

A different model is defined by the prepotential\footnote{Note that both \eqref{F-cubic} and \eqref{F-sqrt} belong to the general class of models with $F$ proportional to $(X^0)^p(X^1)^q$,
where homogeneity of degree two requires $p+q=2$.}
\eq
F = -2i (X^0)^{1/2} (X^1)^{3/2}\,. \label{F-sqrt}
\feq
If we take
\eq
Z^0 = 1\,, \qquad Z^1 = \tau^2\,,
\feq
the holomorphic symplectic vector, K\"ahler potential and scalar metric become respectively
\begin{equation}
v = (1, \tau^2, -i\tau^3, -3i\tau)^t\,, \qquad e^{-\mathcal{K}} = (\tau + \bar\tau)^3\,, \qquad
g_{\tau\bar\tau} = \frac3{(\tau + \bar\tau)^2}\,,
\end{equation}
while the matrix $\cal N$ in \eqref{defN} and the inverse of its imaginary part are given by
\eq
{\cal N} = \frac1{3\tau - \bar\tau}\left(\begin{array}{cc} -2i\tau^3\bar\tau & 3i\tau(\bar\tau - \tau) \\
                3i\tau(\bar\tau - \tau) & -6i\end{array}\right)\,,
\feq
\eq
(\text{Im}\,{\cal N})^{-1} = \frac4{3(\tau + \bar\tau)^3}\left(\begin{array}{cc} -6 & -\frac32(\tau -
                                          \bar\tau)^2 \\
                                          -\frac32(\tau - \bar\tau)^2 & |\tau|^2(\tau^2 - 4\tau\bar\tau + {\bar\tau}^2)
                                          \end{array}\right)\,.
\feq
Finally, the scalar potential reads
\eq
V = -\frac{8g_1}{3(\tau + \bar\tau)}\left[3g_0 + g_1 |\tau|^2\right]\,, \label{pot-sqrt}
\feq
which has a critical point at $\tau^2={\bar\tau}^2=3g_0/g_1$ that corresponds to an AdS vacuum.

\section{Half-BPS rotating near-horizon geometries}
\label{sec:scalars}

\subsection{t$^3$ model}

For the model defined by \eqref{F-cubic}, the flow equation \eqref{flow-Y} boils down to
\begin{equation}
\label{flow-t3}
	\frac{d\tau}{dY} = \frac{(\tau + \bar\tau)\bigl[3\nu + i(\tau - 2\bar\tau)\bigr](\nu + i\tau)}{6(Y- i)
	\bigl(-\nu^2 - i\nu\tau 	+ i\nu\bar\tau - \frac13\tau\bar\tau + \frac13\tau^2 + \frac13\bar\tau^2 \bigr)} ,
\end{equation}
where $\nu\equiv g_0/g_1$.
It turns out that \eqref{flow-t3} can be solved using the assumption
\begin{equation}
\bar\tau = \tau - 2i\mu \,,
\end{equation}
with $\mu$ real and constant.
This leads to
\begin{equation}
\frac{(-3\nu^2 + 6\nu\mu - 4\mu^2 - 2i\mu\tau + \tau^2) d\tau}{(\tau - i\mu)(\tau - i\nu)(\tau +
3i\nu - 4i\mu)} = \frac{dY}{Y-i}\,,
\end{equation}
and thus
\begin{equation}
\label{flow-scalar}
\tau = i\mu + (\mu - \nu)\Bigl(Y\pm\sqrt{Y^2 - 3}\Bigr) \,.
\end{equation}
Note that a different class of half-BPS rotating near-horizon geometries in the $\text{t}^3$ model was 
obtained in \cite{Daniele:2019rpr}.
This is also of the form \eqref{near-hor}, but has $Y=-1/\sqrt{3}$ and is thus clearly not contained in the solutions derived above, where $Y$ was used as a coordinate.

A different way to deal with equation \eqref{flow-t3} is to define the shifted complex scalar
\eq
\hat{\tau}\equiv \tau - i\nu
\feq
and to split the flow equation in its real and imaginary parts,
\eq
\label{flow-complex}
\dfrac{d\hat\tau_R}{dY} = \hat\tau_R \dfrac{Y (\hat\tau_R^2 + 3\hat\tau_C^2) + 2\hat\tau_R
\hat\tau_C}{(1 + Y^2)(\hat\tau_R^2 - 3\hat\tau_C^2)}\,, \qquad
\dfrac{d\hat\tau_C}{dY} = \hat\tau_R \dfrac{\hat\tau_R^2 + 3\hat\tau_C^2 - 2Y \hat\tau_R
\hat\tau_C}{(1 + Y^2)(\hat\tau_R^2 - 3\hat\tau_C^2)}\,,
\feq
where $\hat\tau_R\equiv\mathrm{Re}\hat\tau$ and $\hat\tau_C\equiv\mathrm{Im}\hat\tau$. We can
easily decouple this system by defining the auxiliary function $f(Y)\equiv\hat\tau_C/\hat\tau_R$. In this
way, we get
\eq
f' = \frac{(1 + f^2)(1 - 3Yf)}{(1 + Y^2)(1 - 3f^2)}\,. \label{f'(Y)}
\feq
A more convenient way of dealing with this equation is to suppose the function $f(Y)$ to be invertible and
to rewrite \eqref{f'(Y)} in terms of the inverse $Y(f)$,
\eq
Y' = \frac{(1 + Y^2)(1 - 3f^2)}{(1 + f^2)(1 - 3fY)}\,.
\feq
Since they will guide our steps in the construction of more solutions, we present here the expressions of $f(Y)$ and its inverse in the case of \eqref{flow-scalar},
\eq
f(Y) = \frac1{Y\pm\sqrt{Y^2 - 3}}\,, \qquad Y(f) = \frac{1 + 3f^2}{2f}\,.
\feq
Inspired by these, we define $y(f)\equiv 2f Y(f)$ and suppose a polynomial expansion for it. Then,
it is straightforward to check that $y(f)$ can be at most of degree 2, and the only possible solutions are
\eq
Y(f) = \frac{1 + 3f^2}{2f} \,, \qquad  Y(f) = \frac{1 - f^2}{2f} \,, \qquad  Y(f) = f\,,
\feq
with inverse
\eq
f(Y) = \frac{Y\pm\sqrt{Y^2 - 3}}3\,, \qquad f(Y) = -Y\pm\sqrt{Y^2 + 1}\,, \qquad f(Y) = Y\,. \label{f(Y)}
\feq
The first solution is exactly \eqref{flow-scalar}, as expected. In order to find $\hat\tau_R$ and
$\hat\tau_C$ in the remaining cases, the expression for $f$ has to be plugged into its definition to get 
$\hat\tau_C$ as a function of $\hat\tau_R$ (or viceversa), and with this relation one can try to solve the 
system \eqref{flow-complex}. For the last two cases of \eqref{f(Y)}, this leads respectively to
\eq
\label{flow-scalar-b}
\tau = \mu\frac{\sqrt{Y^2 + 1}\mp 2Y}{\sqrt{Y^2 + 1}}\left(\pm Y + \sqrt{Y^2 + 1}\pm i\right) + i\nu\,,
\feq
\eq
\label{flow-scalar-FP}
\tau = \frac{\mu}{\sqrt{3Y^2 - 1}}\left(1 + iY\right) + i\nu\,.
\feq
For the solution~\eqref{flow-scalar-FP}, the relation~\eqref{Y} between the coordinates $Y$ and $\xi$
boils down to
\eq
\label{Y-FP}
\mu = 8g_1^2 e^{-\xi}\sqrt{3Y^2 - 1}\,.
\feq
Plugging this into \eqref{flow-scalar-FP} gives
\eq
\tau = 8g_1^2 e^{-\xi}\left(1 + iY\right) + i\nu\,, \label{tau(xi,Y)}
\feq
which, contrary to \eqref{flow-scalar-FP}, is well-defined also for $Y^2=1/3$.
For $Y=-1/\sqrt3$ this corresponds exactly to the near-horizon solution costructed
in \cite{Daniele:2019rpr}, cf.~(4.13) and (4.14) of \cite{Daniele:2019rpr}. Eliminating $Y$ in \eqref{tau(xi,Y)}
one gets
\eq
\label{flow-scalar-FP-2}
\tau = 8g_1^2 e^{-\xi}\left(1\pm\frac i{\sqrt3}\sqrt{1 + e^{2\xi}\frac{\mu^2}{64g_1^4}}\right) + i\nu\,,
\feq
and thus the solution in \cite{Daniele:2019rpr} is retrieved for $\mu=0$. It is worth noting that when the scalar is written in terms of $Y$, $\mu=0$ seems not to be an allowed value since, from the expression
for the K\"ahler potential, $\mathrm{Re}\tau > 0$. The origin of this discrepancy resides in the fact that 
\eqref{Y-FP} implies $Y=\pm 1/\sqrt3$ for $\mu=0$.

\subsection{Square root model}

For the model \eqref{F-sqrt}, we define a rescaled scalar field $z$ by
\eq
z = \left(\frac{g_1}{3g_0}\right)^{1/2}\tau\,,
\feq
such that the AdS critical point of the potential \eqref{pot-sqrt} is at $z=1$. Then the flow equation
\eqref{flow-Y} becomes
\eq
\frac{dz}{dY} = \frac{\left(\frac13 + z^2\right)\left(-1 - {\bar z}^2 + 2z\bar z\right)\left(z + \bar z
                        \right)}{2(Y - i)\left[\frac13 - 2z\bar z + z\bar z (z\bar z - z^2 - {\bar z}^2)\right]}\,.
\feq
Solutions for the square root prepotential are actually already known, since the model \eqref{F-sqrt} with 
purely electric gauging is equivalent to the t$^3$ model with mixed gauging considered 
in \cite{Hristov:2018spe}. In fact, one can rotate the t$^3$ model into \eqref{F-sqrt} by a symplectic
transformation with
\begin{equation}
\label{s-matrix}
\mathcal{S} =
\begin{pmatrix}
0 & 0 & -\alpha^{-3} & 0 \\
0 & \alpha & 0 & 0 \\
\alpha^3 & 0 & 0 & 0 \\
0 & 0 & 0 & \alpha^{-1} \\
\end{pmatrix}\,,
\end{equation}
where $\alpha\in\mathbb{R}\!\setminus\! 0$. Notice that $\mathcal{S}$ is fixed by the
following constraints: First of all, it must be symplectic,
\begin{equation}
\mathcal{S}^t\Omega\mathcal{S} = \Omega\,,
\end{equation}
with $\Omega$ given in \eqref{omega}. Moreover, the symplectic sections \eqref{sympl-vec} must satisfy the constraint \eqref{sympconst}, and finally the prepotentials which describe the two models, \eqref{F-cubic} and \eqref{F-sqrt}, must be proportional, $F_\text{sqrt}\propto F_{\text{t}^3}$. It turns out that, in the present case, the correct relation is $F_\text{sqrt}=\pm 2F_{\text{t}^3}$. Without loss of generality we take the upper sign\footnote{Cf.~the comment further below on the case of the lower sign.}, which leads to~\eqref{s-matrix}.

The action of the symplectic rotation~\eqref{s-matrix} on a generic vector of gauge couplings
$\mathcal{G} = (g^0,g^1,g_0,g_1)^t$ is given by
\begin{equation}
\mathcal{G'} = \mathcal{S} \mathcal{G} =
\begin{pmatrix}
-g_0/\alpha^3 \\
g^1 \alpha \\
g^0 \alpha^3 \\
g_1/\alpha
\end{pmatrix},
\end{equation}
and thus the mixed gauging of \cite{Hristov:2018spe} ($g^1=g_0=0$) is rotated into the
purely electric one considered here. Notice also that purely magnetic or purely electric gaugings are
always rotated into mixed ones. Conversely, a mixed gauging may be rotated either into a pure or a
mixed one, depending on which are the nonvanishing FI-parameters. For instance, $g^0=g_1=0$ is 
transformed into a purely magnetic gauging, while $g^0=g_0=0$
is rotated into a mixed one (with the same non-zero coupling constants).

Instead of $\mathcal S$, we can equally choose
\begin{equation}
\mathcal{S}' =
\begin{pmatrix}
-\beta^3 & 0 & 0 & 0 \\
0 & 0 & 0 & -\frac{1}{3\beta} \\
0 & 0 & -\frac{1}{\beta^3} & 0 \\
0 & 3\beta & 0 & 0 \\
\end{pmatrix}
\end{equation}
($\beta\in\mathbb{R}\!\setminus\! 0$), which has the same effect, but leads to $F_\text{sqrt}=-2F_{\text{t}^3}$.

\section{The first solution}
\label{sec:first-solution}

In this section we construct the near-horizon black hole solution associated to the scalar
\eqref{flow-scalar}. Introducing the new coordinate $p\equiv Y\pm\sqrt{Y^2-3}$ and shifting 
$\mu\to\mu+\nu$ for convenience, the scalar field takes the form
\begin{equation}
\tau = i(\mu + \nu) + \mu p\,, \label{tau-first-sol}
\end{equation}
while the metric is given by~\eqref{near-hor},
\begin{equation}
\label{near-hor-p}
\begin{split}
ds^2 & = \frac{p (9+p^2)\Delta}8\biggl(-r^2 dt^2 + \frac{dr^2}{r^2}\biggr) + \frac{p^2 (9+p^2)^2
\Delta^2 - 1024K}{8 p (9+p^2)\Delta} (d\phi + r dt)^2 \\
     & \quad +\frac92\frac{p^3 (9+p^2)\Delta^3}{p^2 (9+p^2)^2\Delta^2 - 1024K} dp^2\,,
\end{split}
\end{equation}
and the gauge potentials follow from~\eqref{fluxes-1/2BPS},
\begin{subequations}\label{gauge-pot-first}
\begin{align}
A^0 & = -\frac{24 g_1\sqrt K}{\mu^2 p^2 (9+p^2)} (d\phi + r dt)\,, \\
A^1 & = \frac{8 g_1\sqrt K\left[3\nu + \mu (3+p^2)\right]}{\mu^2 p^2 (9+p^2)} (d\phi + r dt)\,,
\end{align}
\end{subequations}
where we introduced the constant $\Delta\equiv\mu/g_1^2$.

In order to have the correct signature, we need $p\Delta>0$ and $p^2(9+p^2)^2\Delta^2-1024K>0$\footnote{Notice that $p\Delta>0$ also follows from the constraint $\text{Re}\tau>0$, cf.~the expresssion \eqref{vKt^3} for the K\"ahler potential.}. Without loss of
generality we shall take $p>0$, $\Delta>0$, and thus $\mu>0$. Since $K$ and $p\Delta$ are both
positive, the second constraint reduces to $f_1(p)\equiv p(9+p^2)\Delta-32\sqrt K>0$. In our range of 
parameters the cubic polynomial $f_1$ is characterized by a negative discriminant and has therefore only
one real root $\bar p$ which represents a coordinate singularity. Since the polynomial is positive for 
$p>\bar p$, our spacetime is defined in the interval $p\in(\bar{p},+\infty)$.
Inspection of the scalar curvature shows the presence of a curvature singularity at $p=0$, which
lies outside the allowed domain.
The range of the periodic angular coordinate $\phi$ is fixed by imposing the absence of conical
singularities. The induced metric on surfaces of constant $t,r$ can be written as
\begin{equation}
ds^2\bigl|_{t,r} = \frac{f_1 (f_1 + 64\sqrt K)}{8 p (9+p^2)\Delta} d\phi^2 + \frac{9 p^3 (9+p^2)\Delta^3}
{2 f_1 (f_1 + 64\sqrt K)} dp^2\,. \label{metr-ind-tr}
\end{equation}
Close to the singularity $p\to\bar p$ this becomes
\eq
ds^2\bigl|_{t,r}\to\frac{3\bar p^2\Delta}{3 + \bar p^2}\left[d\rho^2 + \frac{(3 + \bar p^2)^2}{4\bar p^2} 
\rho^2d\phi^2\right]\,,
\feq
where $\rho^2\equiv p-\bar p$. A conical singularity in $p=\bar p$ ($\rho=0$) is thus avoided if
we identify
\eq
\phi\sim\phi + \frac{4\pi\bar p}{3 + \bar p^2}\,.
\feq

\section{The second solution}
\label{sec:second-solution}

Here we shall analyze the spacetime generated by the solution \eqref{flow-scalar-b} to the flow 
equation. In doing this we follow the path traced in the previous section, to which we refer for further details.

We start defining $p\equiv Y\pm\sqrt{Y^2+1}$, which, for consistency, implies $p>0$ when the plus
sign is taken and $p<0$ otherwise. Without loss of generality, in what follows we shall restrict to the first 
case. In terms of $p$ the scalar field becomes
\eq
\label{2nd-scalar}
\tau = \mu\frac{3 - p^2}{1 + p^2}(p+i) + i\nu\,,
\feq
the metric \eqref{near-hor} reads
\begin{equation}
\label{2nd-metric}
\begin{split}
ds^2 = \ &\frac{p (3-p^2)\Delta}8\left(-r^2 dt^2 + \frac{dr^2}{r^2}\right) + \frac{p^2 (3-p^2)^2\Delta^2
- 1024K}{8p (3-p^2)\Delta} (d\phi + r dt)^2 \\
         & + \frac92\frac{p^3 (3-p^2)\Delta^3}{p^2 (3-p^2)^2\Delta^2 - 1024K} dp^2\,,
\end{split}
\end{equation}
and the gauge potentials are given by
\begin{subequations} \label{2nd-gauge-pot}
\begin{align}
A^0 & = \frac{8 g_1\sqrt K}{\mu^2}\frac{1 - 3p^2}{p^2 (3 - p^2)^2}(d\phi + r dt)\,, \\
A^1 & = -\frac{8 g_1\sqrt K}{\mu^2}\frac{\nu (1 - 3p^2) + \mu (1 - p^2)(3 - p^2)}{p^2 (3 - p^2)^2}
(d\phi + r dt)\,,
\end{align}
\end{subequations}
where, again, $\Delta\equiv\mu/g_1^2$.

The range of the coordinate $p$ is defined by $p(3-p^2)\Delta>0$ and $p^2 (3-p^2)^2\Delta^2
-1024K>0$\footnote{Also here the first constraint is equivalent to $\text{Re}\tau>0$.}, hence
$f_2(p)\equiv p(3-p^2)\Delta-32\sqrt K>0$. In this case the analysis of the polynomial $f_2$ is more 
complicated since the discriminant does not have a definite sign. For $\Delta<0$,
$f_2$ has a local maximum at $p=p_{\text M}=-1$, where $f_2(p_{\text M})=-32\sqrt K+
2|\Delta|$, and a local minimum at $p=p_{\text m}=1$, where $f_2(p_{\text m})=-32\sqrt K
-2|\Delta|<0$\footnote{For $\Delta>0$, $p=1$ is a maximum and $p=-1$ a minimum.}.
The polynomial $f_2$ has one root for $|\Delta|<16\sqrt K$, one single and one double root when the 
equality holds and three distinct zeroes for $|\Delta|>16\sqrt K$.

\begin{figure}[h]
\centering
\includegraphics[scale=0.37]{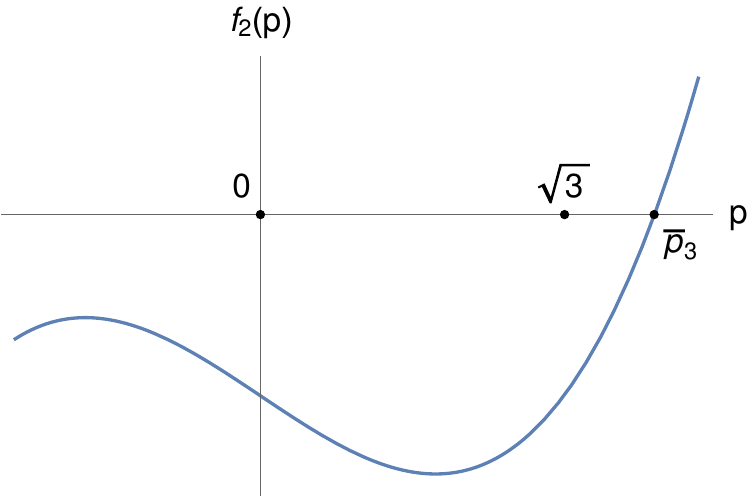} \quad
\includegraphics[scale=0.37]{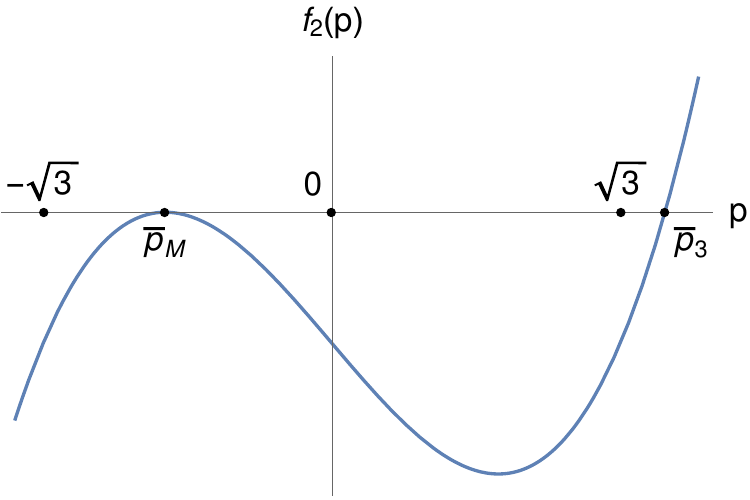} \quad
\includegraphics[scale=0.37]{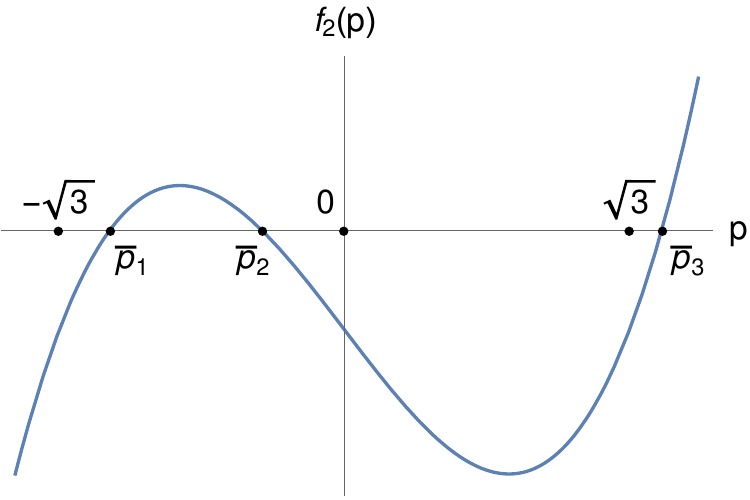}
\caption{Graph of $f_2$ for: (a) $0>\Delta>-16\sqrt K$, (b) $\Delta=-16\sqrt K$, (c) $\Delta<-16\sqrt K$. 
Switching the sign of $\Delta$ produces a reflection of $f_2$ over the vertical axis.}
\label{func-f2}
\end{figure}

From the possible graphs of $f_2$, shown in figure \ref{func-f2} for negative values of $\Delta$, we
derive the following ranges of definition of $p$:
\begin{itemize}
\setlength\itemsep{0em}
\item $\Delta<0$: $p\in(\bar{p}_3,+\infty)$\,,
\item $0<\Delta\leq16\sqrt K$: no interval\,,
\item $\Delta>16\sqrt K$: $p\in(\bar{p}_2,\bar{p}_1)$\footnote{Here the $\bar p$'s have opposite
sign w.r.t.~the ones in figure \ref{func-f2}.}.
\end{itemize}
Notice that $\Delta$ determines the topology of the horizon, which is noncompact for $\Delta<0$
and compact for $\Delta>16\sqrt K$.
Finally, the scalar curvature diverges in $p=0$ and $p=\pm\sqrt3$. Since $f_2(0)=f_2(\pm\sqrt3)<0$,
these points are located outside the allowed regions.

\begin{figure}[h]
\centering
\includegraphics[scale=0.3]{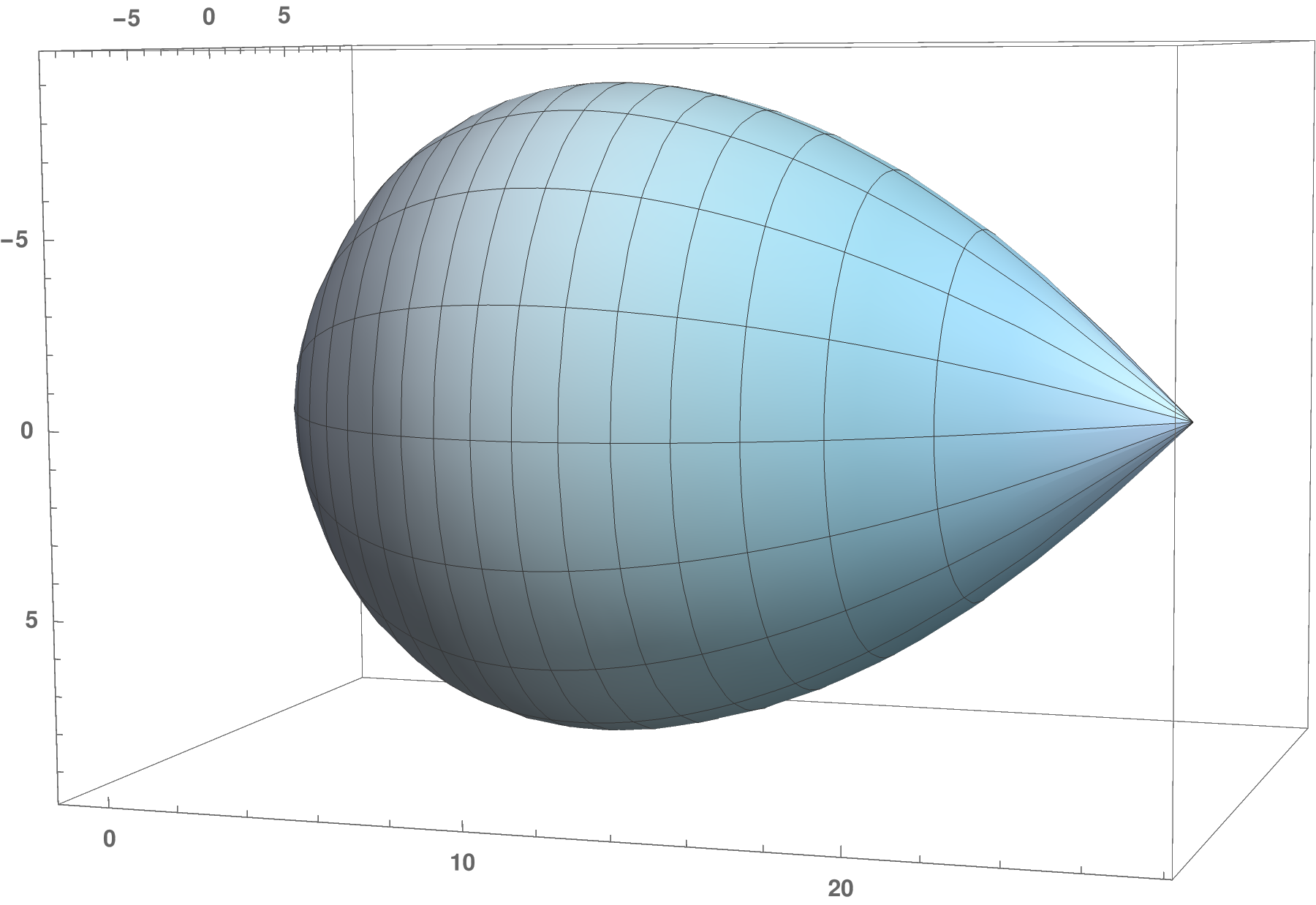}
\caption{Embedding diagram of the horizon of the black hole~\eqref{2nd-metric} for $\Delta=86$ and $K=10$.}
\label{embed}
\end{figure}

In order to fix the periodicity of the angular coordinate $\phi$ we restrict again to surfaces of constant
$t$ and $r$. Close to a generic root $\bar p$ of $f_2$ the metric approaches
\begin{equation}
ds^2\bigl|_{t,r}\to\frac{3\bar p^2 |\Delta|}{|1 - \bar p^2|}\left[d\rho^2 + \frac{(1 - \bar p^2)^2}
{4\bar p^2}\rho^2 d\phi^2\right]\,, \label{ds^2|_tr-2nd}
\end{equation}
where again $\rho^2\equiv|p-\bar p|$. \eqref{ds^2|_tr-2nd} is free from conical singularities
in $p=\bar p$ if
\eq
\phi\sim\phi + \frac{4\pi\bar p}{|1 - \bar p^2|}\,.
\feq
Let us now consider the case of a compact horizon, i.e., $\Delta>16\sqrt K$, $p\in(\bar p_2,\bar p_1)$.
Since $\bar p_2$ lies at the left of the local maximum in $p=1$ and $\bar p_1$ on the right,
we have $\bar p_2^2<1<\bar p_1^2$. Requiring the periodicities of $\phi$ in $\bar p_1$ and $\bar p_2$
to be equal gives thus
\eq
\frac{\bar p_1}{\bar p_1^2 - 1} = \frac{\bar p_2}{1 - \bar p_2^2}\quad\Rightarrow\quad\bar p_1
= \frac1{\bar p_2}\quad (\lor\,\bar p_1 = -\bar p_2)\,,
\feq
where the last case is excluded since we assumed $p>0$. We can then write
\eq
f_2(p) = p(3 - p^2)\Delta - 32\sqrt K = -\Delta\left(p - \frac1{\bar p_2}\right)\!(p - \bar p_2)
(p - \bar p_3)\,.
\feq
Comparing the coefficients of the various powers one obtains $\bar p_2=1$ and therefore also $\bar p_1=1$. One can thus remove only one of the two conical singularities located at the poles, and it is not possible to have a horizon that is smooth in both $\bar p_1$ and $\bar p_2$. The remaining singularity can be thought of as being created by a semi-infinite cosmological string. The horizon geometry is visualized by means of the embedding diagram in figure~\ref{embed}.

Alternatively, one can keep both singularities and tune the parameters of the metric such that the horizon becomes a specific kind of orbifold\footnote{We recall that an $n$-dimensional orbifold is a topological space locally modelled on $\RR^n/\Gamma$, where $\Gamma$ are finite groups.}, the weighted projective space $\Sigma=\text{W}\CC\text{P}^1_{[n_1,n_2]}$, also called spindle (see~\cite{Ferrero:2020laf,Ferrero:2020twa,Hosseini:2021fge} for a list of recent references on this topic). This space is topologically a 2-sphere, but with conical singularities at the poles, characterized by deficit angles $2\pi(1-1/n_{1,2})$, with $n_{1,2}$ two coprime positive integers. Locally, the poles are modelled on $\RR^2/\ZZ_{n_1}$ and $\RR^2/\ZZ_{n_2}$. Fixing appropriately the constant $\sqrt{K}/\Delta$ and the period of $\phi$, the two-dimensional metric spanned by $p$ and $\phi$ becomes a smooth metric on the orbifold $\Sigma$.

\subsection{Uplifting to $D=5$}

Sometimes a better understanding of a given solution may be gained by taking advantage of the correlation between four and five-dimensional supergravity theories and studying the counterpart of the  initial background. In this subsection we shall follow this strategy.

Let us consider the black hole solution \eqref{2nd-scalar}-\eqref{2nd-gauge-pot} and uplift it to $N=2$, $D=5$ FI-gauged supergravity through the $r$-map (cf.~appendix \ref{app:r-map}). Since the starting point is the four-dimensional t$^3$ model, the result will solve the equations of motion of five-dimensional pure gauged supergravity, whose bosonic sector consists of an Einstein-Maxwell theory with Chern-Simons term and cosmological constant. The oxidized solution reads
\begin{equation}
\begin{aligned}
ds^2 &= \frac{1+x}{8g_1^2} \biggl( -r^2 dt^2 + \frac{dr^2}{r^2} \biggr) + \frac{1}{8g_1^2} \frac{(1+x) \bigl[x (3-x)^2 - 1024K\bigr]}{x (3-x)^2} (d\phi + r dt)^2 \\
     & \quad\, + \frac{9}{8g_1^1} \frac{1+x}{x (3-x)^2 - 1024K} \, dx^2 + \frac{x (3-x)^2}{(1+x)^2} \biggl[ d\psi + \frac{8\sqrt{2K}}{g_1} \frac{1-3x}{x (3-x)^2} (d\phi + r dt) \biggr]^2 \,, \\[0.5em]
A &= \sqrt{3} \, \frac{3-x}{1+x} \, d\psi - \frac{8\sqrt{6K}}{g_1} \frac{1}{1+x} (d\phi + r dt) \,,
\end{aligned}
\end{equation}
where we defined the new coordinate $x=p^2$ and rescaled $K\mapsto \Delta^2 K$ in order to get rid of $\mu$.

Even though this expression may seem not very instructive, setting $K=0$ it is possible to retrieve the spindle solution of~\cite{Ferrero:2020laf}, namely equations~(3)-(5), with $a=1$, a limiting case in which the horizon collapses to a circle. Despite the analogy between the four- and five-dimensional solutions, the two spindles have quite a different origin. Indeed, in the uplifted solution with vanishing $K$, $\phi$, which played the role of horizon coordinate in $D=4$, now parametrizes an AdS$_3$ space (cf.~\cite{Ferrero:2020laf}), while the Kaluza--Klein coordinate $\psi$ gives birth, along with $x$, to the spindle.

\section{Black hole extension}
\label{sec:third-solution}

In what follows we shall construct and analyze the extension of the solution associated to
\eqref{flow-scalar-FP-2} to the whole spacetime outside the black hole. To this end, it is convenient to 
rewrite the solution in a different coordinate system, which allows, by comparing to
\cite{Daniele:2019rpr}, to extend the metric to the region far from the horizon.

The metric is given by \eqref{near-hor}, where
\eq
Y(\xi) = \frac1{\sqrt3}\sqrt{1 + \frac{\mu^2}{64 g_1^4} e^{2\xi}}\,,
\feq
which follows from \eqref{Y-FP}. The potentials related to the fluxes \eqref{fluxes-1/2BPS} are
\begin{subequations}
\begin{align}
A^0 & = -\frac{\sqrt K e^{2\xi}}{16 g_1^3}(d\phi + r dt)\,, \\
A^1 & = -\frac{\sqrt K e^{2\xi}}{16 g_1^3}\left(\mp 8 g_1^2 e^{-\xi} Y(\xi) - \frac{g_0}{g_1}\right)
(d\phi + r dt)\,.
\end{align}
\end{subequations}
With the change of coordinates
\begin{equation}
e^{-\xi} = \sqrt K\coth\tilde x\,, \qquad\phi = \sqrt3 y\,, \qquad  t = \frac T{2\sqrt K}\,,
\end{equation}
we can recast the scalar, metric and gauge potentials in a form that closely resembles the one in
section 4.1 of \cite{Daniele:2019rpr},
\begin{equation}
\tau = R(\tilde x) \bigl(\sqrt3\pm i F(\tilde x)\bigr) + i\frac{g_0}{g_1}\,,
\end{equation}
\begin{equation}
ds^2 = -\frac{8 g_1^2}{\sqrt3 R(\tilde x)}\left(r dT + \frac3{4g_1^2}\partial_{\tilde x} R(\tilde x) dy
\right)^2 + \frac{\sqrt3 R(\tilde x)}{2 g_1^2}\left(\frac{dr^2}{r^2} + \frac{3 (dx^2 + dy^2)}{\sinh^2\!
\tilde x}\right),
\end{equation}
\begin{subequations}
\begin{align}
A^0 & = -\frac{2 g_1}{3 R(\tilde x)^2}\left(r dT + \frac3{4 g_1^2}\partial_{\tilde x} R(\tilde x) dy\right)\,, \\
A^1 & = -\frac{2 g_1}{3 R(\tilde x)^2}\left(\mp R(\tilde x) F(\tilde x) - \frac{g_0}{g_1}\right)\left(r dT + 
\frac3{4 g_1^2}\partial_{\tilde x} R(\tilde x) dy\right)\pm\frac{\coth\tilde x}{2 g_1} F(\tilde x) dy\,,
\end{align}
\end{subequations}
where
\eq
R(\tilde x) = \Xi_1\coth\tilde x\,, \qquad\Xi_1 = 8 g_1^2\sqrt{\frac K3}\,, \qquad
\frac{d\tilde x}{dx} = \mp F(\tilde x)\equiv\mp\sqrt{1 + \frac{\mu^2}{3 R(\tilde x)^2}}\,.
\feq
Inspired by \cite{Daniele:2019rpr}, we make the following ansatz for the full black hole extension
($\alpha$ and $\beta$ are real constants):
\begin{equation}
\tau = \sqrt3\sqrt{\alpha r + \beta + R^2}\pm i R F + i\frac{g_0}{g_1}\,, \label{tau-fullbh}
\end{equation}
\begin{equation}
\begin{split}
ds^2 &= -\frac{8 g_1^2}{\sqrt3\sqrt{\alpha r + \beta + R^2}}\left(r dT + \frac3{4 g_1^2}
\partial_{\tilde x} R dy\right)^2 \\
& \quad +\frac{\sqrt3\sqrt{\alpha r + \beta + R^2}}{2 g_1^2}\left(\frac{dr^2}{r^2} + \frac{3 (dx^2 + dy^2)}
{\sinh^2\!\tilde x}\right),
\end{split} \label{metr-fullbh}
\end{equation}
\begin{subequations}
\begin{align}
A^0 & = -\frac{2 g_1}{3(\alpha r + \beta + R^2)}\left(r dT + \frac3{4 g_1^2}\partial_{\tilde x} R dy\right), \\
A^1 & = -\frac{2 g_1}{3(\alpha r + \beta + R^2)}\left(\mp R F - \frac{g_0}{g_1}\right)\left(r dT + 
\frac3{4 g_1^2}\partial_{\tilde x} R dy\right)\pm\frac{\coth\tilde x}{2 g_1} F dy\,,
\end{align} \label{A-fullbh}
\end{subequations}
which we have shown to satisfy all the BPS equations of section \ref{subsec:BPSeqns}.
Eqns.~\eqref{tau-fullbh}-\eqref{A-fullbh} represent a generalization of the black hole solution
constructed in section~4.1 of~\cite{Daniele:2019rpr} with $\Xi_2=0$. If the free parameter $\mu$
vanishes, our solution boils down to the one in~\cite{Daniele:2019rpr} with $\Xi_2=0$.

The spacetime~\eqref{metr-fullbh} has a horizon for $r=0$, with induced metric
\begin{equation}
ds^2\bigl|_{t,r} = \frac{3\sqrt3}{2 g_1^2\sinh^2\!\tilde x}\left[\sqrt{\beta + R^2} dx^2 +
\frac{\beta + \Xi_1^2}{\sqrt{\beta + R^2}} dy^2\right]\,.
\end{equation}
As expected, the metric \eqref{metr-fullbh} is not asymptotically AdS$_4$ since the scalar potential in \eqref{scalar-pot-t^3} has no critical points. A detailed physical discussion of the supersymmetric black hole given by~\eqref{tau-fullbh}-\eqref{A-fullbh} will be presented elsewhere.

\subsection{Uplifting to $D=5$}

In order to gain a deeper insight into this solution, once again we apply the $r$-map and uplift~\eqref{tau-fullbh}-\eqref{A-fullbh} to five dimensions. The five-dimensional background is
\begin{equation}
\begin{aligned}
ds^2 &= -\frac{8 g_1^2}{3f} \left(r dT + \frac3{4 g_1^2} \partial_{\tilde x} R dy\right)^2 +\frac{1}{2 g_1^2} \left(\frac{dr^2}{r^2} + \frac{3(dx^2 + dy^2)}{\sinh^2\!\tilde{x}}\right) \\
& \quad\ + 3f \left[ dz - \frac{\sqrt{8} g_1}{3f} \left(r dT + \frac3{4 g_1^2} \partial_{\tilde x} R dy\right) \right]^2 \,, \\
A &= \pm F \coth\tilde{x} \left(\Xi_1 dz + \frac{dy}{\sqrt{2} g_1}\right) \,,
\end{aligned}
\end{equation}
where we defined
\begin{equation}
f(r,\tilde{x}) = \alpha r + \beta + R(\tilde{x})^2 \,.
\end{equation}
Although written in an obscure form, this is actually a product space between an extremal BTZ black hole
and the hyperbolic plane. To see this, start by shifting the $y$ coordinate as $\hat y=y+\sqrt2 g_1\Xi_1 z$, 
which yields a simplified expression for the metric and the gauge field,
\begin{equation}
\begin{aligned}
ds^2 &= -\frac{8 g_1^2}{3f_0} r^2 dT^2 + \frac{dr^2}{2 g_1^2 r^2} + \left(\sqrt{3f_0} dz - \frac{\sqrt8
g_1}{\sqrt{3f_0}} r dT\right)^2 + \frac3{2 g_1^2}\frac{dx^2 + d\hat{y}^2}{\sinh^2\!\tilde x}\,, \\
A &=\pm\frac F{\sqrt2 g_1}\coth\tilde x d\hat y\,,
\end{aligned}
\end{equation}
with
\begin{equation}
f_0(r) = \alpha r + \beta + \Xi_1^2\,.
\end{equation}
The canonical two-dimensional hyperbolic metric appears explicitely defining
\begin{equation}
\lambda\cosh\theta = \pm F\coth\tilde x\,, \qquad\phi = \hat y\sqrt{1 + \frac{\mu^2}{3\Xi_1^2}}\,,
\end{equation}
while the BTZ metric can be retrieved by the coordinate change
\begin{equation}
t = \frac{2g_1^2}{\sqrt3\alpha} T\,,\qquad\rho^2 = f_0(r)\,,\qquad\psi = -\sqrt{\frac32} g_1 z + 
\frac{2g_1^2}{\sqrt3\alpha} T\,.
\end{equation}
The result is
\begin{equation}\label{BTZxH2}
\begin{aligned}
ds^2 &= \frac2{g_1^2}\left[-N^2 dt^2 + \frac{dr^2}{N^2} + \rho^2 (d\psi + N^\psi dt)^2\right] +
\frac3{2 g_1^2} (d\theta^2 + \sinh^2\!\theta d\phi^2)\,, \\
A &= \frac1{\sqrt2 g_1}\cosh\theta d\phi\,,
\end{aligned}
\end{equation}
where
\begin{equation}
N = \rho - \frac{\beta + \Xi_1^2}\rho\,,\qquad N^\psi = -\frac{\beta + \Xi_1^2}{\rho^2}\,,
\end{equation}
which are, respectively, the lapse and shift functions that define an extremal BTZ black hole with mass 
$M=2(\beta+\Xi_1^2)$. It is worth noting that the solution is described by the only free parameter $M$, 
contrary to the three constants characterizing the four-dimensional system\footnote{If $\alpha$ is
nonvanishing, its value in \eqref{tau-fullbh}-\eqref{A-fullbh} can be set equal to 1 by rescaling $r$ and 
$T$.}. The behaviour we observe is somewhat unexpected: the full black hole solution~\eqref{tau-fullbh}-\eqref{A-fullbh} reduces to the near-horizon of a black string with momentum along the string once uplifted to five dimensions. Notice in this context that the extremal BTZ black hole
preserves half of the supersymmetries of $(1,1)$ AdS supergravity in three
dimensions~\cite{Coussaert:1993jp} and that the whole five-dimensional background can be proved to be 1/2-BPS, as expected for a near-horizon geometry.

\section{Analytical continuation to NUT black holes}
\label{sec:nut}

It is possible to analytically continue the near-horizon metric of an extremal rotating black hole to obtain
a spacetime with NUT charge. In the following, we apply this procedure to a standard example, namely the 
extreme Kerr throat geometry, and then to the rotating near-horizon solutions described in the previous 
sections.

\subsection{From the extremal Kerr black hole to a NUT spacetime}

Let us consider the near-horizon metric of the extremal Kerr black hole~\cite{Bardeen:1999px},
\begin{equation}
\label{kerr}
ds^2 = \frac{1 + \cos^2\!\theta}2\biggl(-\frac{r^2}{r_0^2} dt^2 + \frac{r_0^2}{r^2} dr^2 + r_0^2 d\theta^2 
\biggr) + \frac{2 r_0^2\sin^2\!\theta}{1 + \cos^2\!\theta} \biggl(d\phi + \frac r{r_0^2} dt\biggr)^2\,,
\end{equation}
where the constant $r_0$ is related to the rotation parameter by $r_0^2=2a^2$.

We perform a double Wick rotation on the coordinates $t$ and $\phi$ by
$t\to i\varphi$, $\phi\to i\tau$. Furthermore, we introduce the new radial coordinate
$\rho = N\cos\theta$ and rename $a\to N$, where $N$ is interpreted as the NUT parameter.
It is worth noting that after the analytic continuation the coordinate $r$ should not be considered radial 
anymore.
Finally, the rescaling $r\to 2N^2 r$ and $\tau\to\tau/2N$ brings the metric \eqref{kerr} to the form
\begin{equation}
\label{nut-metric}
ds^2 = -f(\rho) (d\tau + 2 N r d\varphi)^2 + \frac{d\rho^2}{f(\rho)} + (N^2 + \rho^2)\biggl(r^2
d\varphi^2 + \frac{dr^2}{r^2}\biggr)\,,
\end{equation}
where the radial function is defined as
\begin{equation}
f(\rho) = \frac{N^2 - \rho^2}{N^2 + \rho^2}\,.
\end{equation}
Equation \eqref{nut-metric} is the hyperbolic NUT spacetime \cite{Taub:1950ez,Newman:1963yy}
with parameter $N$.

\subsection{NUT black holes}

Since the supersymmetric near-horizon geometry \eqref{near-hor} has the same form as \eqref{kerr},
one can try to apply the analytical continuation outlined above also in that case.
We shall do this for the two particular examples \eqref{near-hor-p} and \eqref{2nd-metric} to obtain
black holes with NUT charge.

By Wick-rotating the coordinates $t\to i\varphi$, $\phi\to i\tau$ in \eqref{near-hor-p}, we find
\begin{equation}
\label{nut-bh}
\begin{split}
ds^2 & = -\frac{p^2 (9+p^2)^2\Delta^2 - 1024K}{8 p (9+p^2)\Delta} (d\tau + r d\varphi)^2 \\
     &\quad + \frac92\frac{p^3 (9+p^2)\Delta^3}{p^2 (9+p^2)^2\Delta^2 - 1024K} dp^2 + \frac{p
(9+p^2)\Delta}8\biggl(r^2 d\varphi^2 + \frac{dr^2}{r^2}\biggr)\,,
\end{split}
\end{equation}
where now $p$ has radial character. The very same transformations must be applied to the other fields of
the theory, namely the fluxes and the scalar, and one can verify that the new fields still satisfy the equations 
of motion. Note also that \eqref{nut-bh} is of Petrov-type D. In order to have real gauge potentials and thus
real charges, it is necessary to analytically continue the gauge coupling constant as well, $g\to -ig$.
This implies $g_I\to -ig_I$, and thus \eqref{gauge-pot-first} leads to
\begin{subequations}
\begin{align}
A^0 & = -\frac{24 g_1\sqrt K}{\mu^2 p^2 (9+p^2)}(d\tau + r d\varphi)\,, \\
A^1 & = \frac{8 g_1\sqrt K\left[3\nu + \mu (3+p^2)\right]}{\mu^2 p^2 (9+p^2)}(d\tau + r d\varphi)\,,
\end{align}
\end{subequations}
while the scalar field is still given by \eqref{tau-first-sol}\footnote{The relation between
$\Delta$ and $\mu$ introduced in section \ref{sec:first-solution} is actually
$\Delta=\mu/|g_1^2|$, where the absolute value stems from $|X|^2$ in \eqref{Y}. After taking
$g_1\to -ig_1$ one has thus still $\Delta=\mu/g_1^2$.}.
Note that the Wick rotation of the coupling constant amounts to changing the theory from genuine
supergravity to fake supergravity \cite{Meessen:2009ma}, where an $\mathbb{R}$-symmetry instead
of a $\text{U}(1)$ is gauged.
Introducing the new coordinate $R= p^{1/2}$ and taking the limit $R\to\infty$, the asymptotic metric
takes the simple form of a domain wall,
\begin{equation}
\label{nut-dw}
ds^2 = \frac{R^6}{144}\biggl[-(d\tau + r d\varphi)^2 + r^2 d\varphi^2 + \frac{dr^2}{r^2}\biggr] + dR^2\,,
\end{equation}
where we rescaled the metric by a factor of $18\Delta$. Notice that the world volume of this domain
wall is curved, since the metric in brackets in \eqref{nut-dw} is AdS$_3$, written as a Hopf-like
fibration over H$^2$, cf.~e.g.~\cite{Klemm:2014rda}.

A second black hole with NUT charge can be generated by Wick-rotating the near-horizon geometry
\eqref{2nd-metric}, which leads to
\begin{equation}\label{metr-nut-2}
\begin{split}
ds^2 & = -\frac{p^2 (3 - p^2)^2\Delta^2 - 1024K}{8 p (3 - p^2)\Delta} (d\tau + r d\varphi)^2 \\
     &\quad +\frac92\frac{p^3 (3 - p^2)\Delta^3}{p^2 (3 - p^2)^2\Delta^2 - 1024K} dp^2
+ \frac{p (3 - p^2)\Delta}8\left(r^2 d\varphi^2 + \frac{dr^2}{r^2}\right)\,,
\end{split}
\end{equation}
which is also of Petrov-type D, and $p$ is now a radial coordinate. Sending $g$ to $-ig$, the
gauge potentials \eqref{2nd-gauge-pot} become
\begin{subequations}\label{gauge-pot-nut-2}
\begin{align}
A^0 & = \frac{8 g_1\sqrt K (1 - 3 p^2)}{\mu^2 p^2 (3 - p^2)^2} (d\tau + r d\varphi)\,, \\
A^1 & = -\frac{8 g_1\sqrt K\left[\nu (1 - 3 p^2) + \mu (1 - p^2)(3 - p^2)\right]}{\mu^2 p^2 (3 - p^2)^2} 
(d\tau + r d\varphi)\,,
\end{align}
\end{subequations}
and the scalar field is the same as in \eqref{2nd-scalar}, with $\mu=g_1^2\Delta$.
We have checked explicitely that
\eqref{metr-nut-2}, \eqref{gauge-pot-nut-2}, together with \eqref{2nd-scalar}, satisfy the equations
of motion of fake gauged supergravity.
For $\Delta<0$ and large $p$, the metric \eqref{metr-nut-2} asymptotes again to the domain wall 
\eqref{nut-dw}, where $R=p^{1/2}$.

We conclude this section by noting that solutions \eqref{nut-bh} and \eqref{metr-nut-2}
represent genuine black hole metrics.
They both have a curvature singularity in $p=0$ covered by an event horizon.
It is then appropriate to call them `NUT black holes'.

\appendix

\section{$r$-map}
\label{app:r-map}

$N=2$ gauged supergravity theories in four and five dimensions are closely related by a special
dimensional reduction called $r$-map, which connects the latter to the subclass of four-dimensional
theories with cubic prepotential
\begin{equation}
F = -\frac16 C_{\alpha\beta\gamma} \frac{X^\alpha X^\beta X^\gamma}{X^0}\,.
\end{equation}
Here we shall consider vector multiplets only and focus on an abelian gauging, following the conventions
of \cite{Klemm:2016kxw}\footnote{In order to be consistent with our conventions it is necessary to switch
the sign of $C_{\alpha\beta\gamma}$ and $h^\alpha$ and to take $\varepsilon_{(5)}^{\mu\nu\rho\sigma 
z}=\varepsilon_{(4)}^{\mu\nu\rho\sigma}$.}. The dimensional reduction is performed along an $S^1$ with 
compact coordinate $z$ by means of the Kaluza-Klein ansatz
\begin{equation}
\begin{gathered}
ds_{(5)}^2 = e^\frac{\phi}{\sqrt3} ds_{(4)}^2 + e^{-\frac{2\phi}{\sqrt3}} (dz + K_\mu dx^\mu)^2\,,\qquad
A^\alpha = B^\alpha (dz + K_\mu dx^\mu) + C^\alpha_\mu dx^\mu\,, \\
z^\alpha = e^{-\frac{\phi}{\sqrt3}} h^\alpha + i B^\alpha\,, \qquad e^\mathcal{K} = \frac18
e^{\sqrt3\phi}\,,\qquad F^I_{\mu\nu} = \frac1{\sqrt2} (dK_{\mu\nu}, dC^\alpha_{\mu\nu})\,,
\end{gathered}
\end{equation}
where all the fields depend only on the four-dimensional coordinates $x^\mu$. We recall that
$\alpha=1,\ldots,n$ and $I=0,\ldots,n$, and the complex scalars $z^\alpha$ are defined through the 
projection $z^\alpha=-iX^\alpha/X^0$. Moreover, one has to impose the constraints
\begin{equation}
g_{(4)} = 3\sqrt2 g_{(5)}\,,\qquad\xi_0 = 0\,,
\end{equation}
where $g_{(4)}$ and $g_{(5)}$ are the gauge couplings in four and five dimensions respectively.

\section*{Acknowledgements}

This work was supported partly by INFN and by MIUR-PRIN contract 2017CC72MK003.

\end{document}